\documentclass[%
 reprint,
 twocolumn,
 nofootinbib,
 amsmath,amssymb,
 aps,
 pre,
]{revtex4-1}

\usepackage{amsthm}

\makeatletter
\newsavebox{\@brx}
\newcommand{\llangle}[1][]{\savebox{\@brx}{\(\m@th{#1\langle}\)}%
  \mathopen{\copy\@brx\kern-0.5\wd\@brx\usebox{\@brx}}}
\newcommand{\rrangle}[1][]{\savebox{\@brx}{\(\m@th{#1\rangle}\)}%
  \mathclose{\copy\@brx\kern-0.5\wd\@brx\usebox{\@brx}}}
\makeatother

\usepackage[usenames,dvipsnames,svgnames]{xcolor}
\usepackage{graphicx}
\usepackage{dcolumn}
\usepackage{bm}
\usepackage[colorlinks=true,
            linkcolor=blue,
            urlcolor=blue,
            citecolor=blue,
            urlbordercolor = white,
            ]{hyperref}
\usepackage[utf8]{inputenc}
\usepackage[english]{babel}

\usepackage[notref,notcite,final]{showkeys}
\usepackage{subfig}
\def\p{{\partial}}
\def\Re{\mathop{\text{Re}}\limits}
\def\Im{\mathop{\text{Im}}\limits}

\newcommand{\grad}{\mathop{\mathrm{grad}}\limits}

\RequirePackage{color}

\begin{document}
\title{Statistical mechanics of stochastic growth phenomena}

\author{Oleg Alekseev}
\author{Mark Mineev-Weinstein}

\affiliation{
International Institute of Physics,
Federal University of Rio Grande do Norte, 59078-970, Natal, Brazil
}

\begin{abstract}

We develop statistical mechanics for stochastic growth processes and apply it to Laplacian growth by using its remarkable connection with a random matrix theory. The Laplacian growth equation is obtained from the variation principle and describes adiabatic (quasistatic) thermodynamic processes in the two-dimensional Dyson gas. By using Einstein's theory of thermodynamic fluctuations we consider transitional probabilities between thermodynamic states, which are in a one-to-one correspondence with simply connected domains occupied by gas. Transitions between these domains are described by the stochastic Laplacian growth equation, while the transitional probabilities coincide with a free-particle propagator on an infinite-dimensional complex manifold with a K\"ahler metric.
\end{abstract}
\maketitle

\section{Introduction}

An important breakthrough occurred in the early 2000's after realizing a rich integrable structure of the Laplacian growth problem~\cite{Mark2000,tau2000}. Remarkable connections of Laplacian growth with integrable hierarchies and random matrices provide ample opportunities to address long-standing problems with novel methods. A particularly important example is a consolidation of the (two-dimensional) 2D Dyson gas theory~\cite{Dyson}, quantum Hall effect~\cite{Chau}, diffusion-limited aggregation, and Laplacian growth within a single framework of normal random matrix ensembles with complex eigenvalues.

Consider the generalization of a Gaussian ensemble of random matrices~\cite{Ginibre,Girko} to the case of normal random $N\times N$ matrices. The partition function is
\begin{equation}\label{M-int}
	\tau=\int dM dM^\dagger d^{-\frac1\hbar \text{Tr}\left[MM^\dagger+V(M)+\bar{V}(M^\dagger)\right]},
\end{equation}
where $\hbar$ is a positive parameter, $V(z)$, $\bar{V}(z)$ are analytic functions, and the overbar denotes complex conjugation. Since $[M, M^\dagger ] = 0$, the normal matrices, $M$ and $M^\dagger$, can be diagonalized simultaneously. Their complex eigenvalues, $z_n=x_n+i y_n$, can be considered as the positions of particles in a 2D plane. Reducing the matrix integral~\eqref{M-int} to the integral over the eigenvalues, and interpreting the Jacobian of the transformation (which takes the form of the Vandermonde determinant) as an effective logarithmic repulsion between the eigenvalues, the partition function $\tau$ represents 2D Coulomb gas in the external potential~\cite{Chau}.

When the size of the matrices $N$ becomes large some new features emerge, and the language of statistical equilibrium thermodynamics provides an adequate description of the partition function~\eqref{M-int}. Different aspects of the $1/N$ expansion of the free-energy and density correlation functions were already discussed in Refs.~\cite{LargeCor,LargeN}. At large $N$, the eigenvalues of normal matrices densely fill the domain $D$ in the complex plane (the support of eigenvalues), and their density steeply drops down at the edge. In the {\it semiclassical limit}, as $\hbar\to0$ with $t_0=\hbar N$ fixed, the leading contribution to the free energy $F=-\lim_{\hbar\to0}\hbar^{2}\log \tau$ tends to the tau function of analytic curves~\cite{Mark2000}.

Remarkably, the growth of an electronic droplet, while increasing the number of electrons, $N\to N+1$, is equivalent to the Laplacian growth problem~\cite{Teo}, which describes the dynamics of the oil-water interface in a thin gap between two parallel plates~\cite{LG-intro}. In a cell, the velocity of viscous fluid obeys Darcy's law, $\mathbf v=-\grad p$ (in scaled units), where the pressure $p(z)$ as a function of $z=x+i y$ satisfies Laplace equation with a sink at infinity, i.e.,\ $\nabla^2p=0$, and  $p(z)=-\log|z|$ as $z\to\infty$. If one wants to neglect the surface tension, $p= 0$ at the interface between two fluids. The kinematic identity requires the normal interface velocity to be equal to the fluid velocity at the interface, $v_n=-\p_n p(\zeta)$, $\zeta\in \p D$, where $\p_n$ stands for the normal derivative at the boundary. It is convenient to introduce the time-dependent conformal map $w(z,t)$ of the domain occupied by oil $D(t)$ to the exterior of the unit disk, $|w|>1$, normalized such that $w(\infty,t)=\infty$ and $w'(\infty,t)=1/r(t)$ is a real positive function of $t$.\footnote{Here and below dot and prime denote partial derivatives with respect to $t$ and $z$.} In terms of the conformal map, the interface velocity reads
\begin{equation}
	v_n(\zeta,t)=|w'(\zeta,t)|,\qquad \zeta\in \p D(t).
\end{equation}
The same equation of motion describes the {\it adiabatic} (quasistatic) growth of a semiclassical electronic droplet in a nonuniform magnetic field, when the number of electrons increases~\cite{Teo}, and the passages between the equilibrium states are only considered---this is a limitation of the equilibrium thermodynamics.

This paper aims to apply common methods of nonequilibrium thermodynamics to study small fluctuations of the external potential $V(z)$ in the partition function~\eqref{M-int}. This branch of thermodynamics, also known as linear (or weakly nonequilibrium) thermodynamics as introduced by Onsager~\cite{Onsager}, allows one to focus on transitions between thermodynamic states represented by different  electronic droplets. The probability of transitions between thermodynamic states is an essential step towards the path-integral formulation of Laplacian growth.

The structure of the paper is as follows. After introducing the partition function of the 2D Dyson gas in the large $N$ limit~\eqref{Z-rho}, we derive the Laplacian growth equation~\eqref{LG} by maximizing the transitional probability between equilibrium states. Afterwards, following Einstein's theory of thermodynamic fluctuations~\cite{Einstein} we determine the probability for occurring fluctuations of the parameter of the matrix ensemble~\eqref{P-l-tt}, which generates {\it stochastic Laplacian growth} of the domain~\eqref{LG-stochastic}. Remarkably, the growth probability is determined by the free-particle propagator on the infinite-dimensional complex manifold with the K\"ahler metric~\eqref{gkl}. Finally, we draw our conclusions and indicate some open problems.

\section{Two-dimensional Dyson gas}\label{DG}

\subsection{The statistical ensemble}

As we only consider the semiclassical fluctuations of the boundary of the electronic droplet, it is convenient to start with an eigenvalue version of the matrix integral~\eqref{M-int} directly. The partition function of $N$ Coulomb charges repelling each other according to the law of 2D electrostatics in the external potential is\footnote{At some particular values of $\beta$ the partition function is the eigenvalue version of normal matrix ensemble~\eqref{M-int} ($\beta$=1) and normal self-dual matrices ($\beta=2$).}
\begin{equation}\label{Z}
	e^{-\beta F}=\int_{-i\infty}^{i\infty } d\mu\int_{\mathbb C^N} e^{\beta(\mu N - E)}\prod_{i=1}^N d^2z_i,
\end{equation}
where $\mu$ is the Lagrange multiplier, which fixes the total number of charges.\footnote{We substitute the delta-function under the partition function $\delta(N-N^*)=\int_{-i \infty}^{i\infty} d\mu e^{\mu (N-N^*)}$, drop the inessential term $\mu N^*$, and rescale $\mu\to \beta\mu$ afterwards.} The energy of the gas equals\footnote{In what follows $log$ denotes a logarithm with the base ``$e$''.}
\begin{equation}\label{E}
	-  E = \sum_{i\neq j}^N\log|z_i-z_j|+ \sum_{i=1}^N U(z_i),
\end{equation}
where $U(z)$ is the external potential.
\begin{figure}[t]
	\centering
	\subfloat[{}\label{droplet3-a}]{
		\includegraphics[width=0.4\columnwidth]{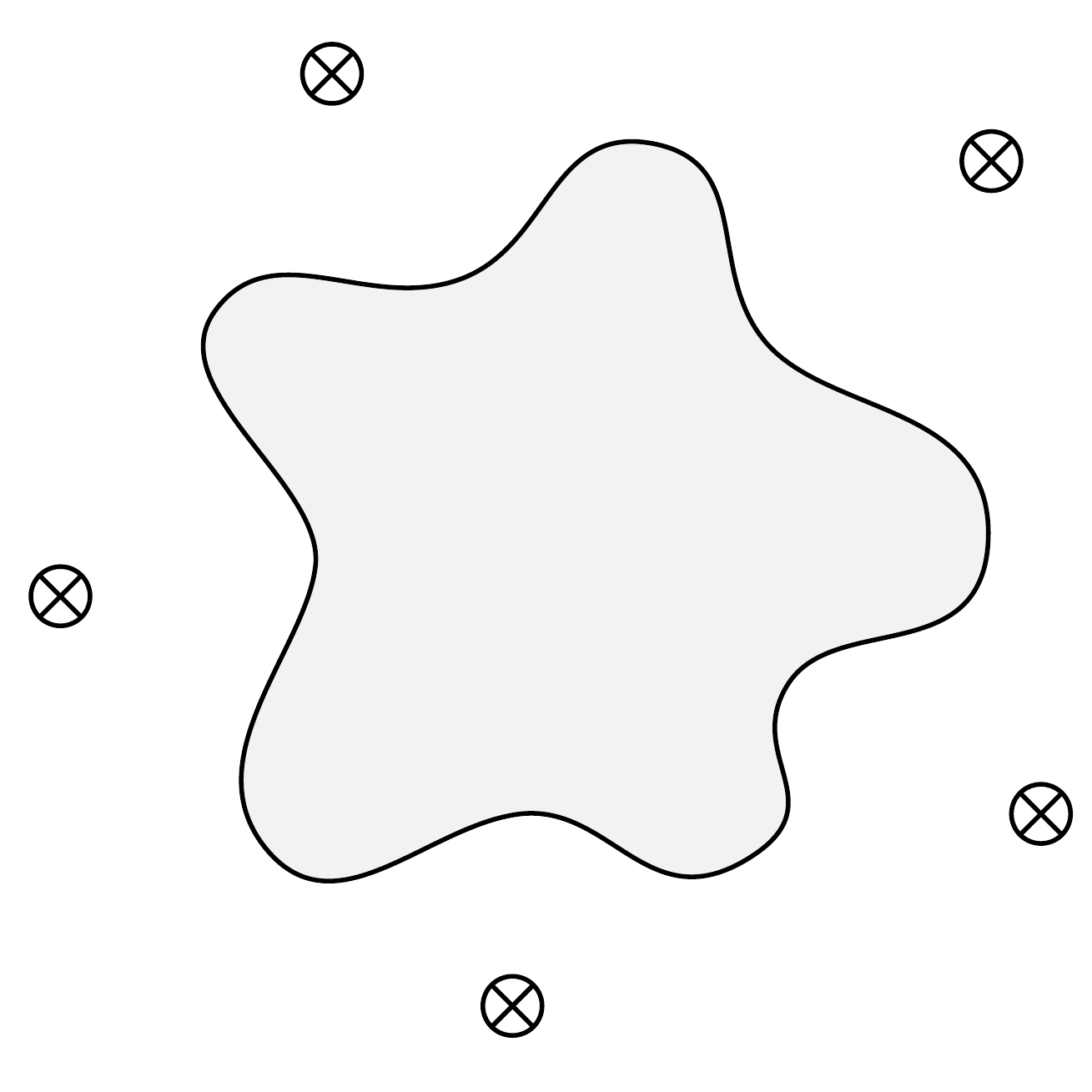}
	}
	\subfloat[{}\label{droplet3-b}]{
		\includegraphics[width=0.4\columnwidth]{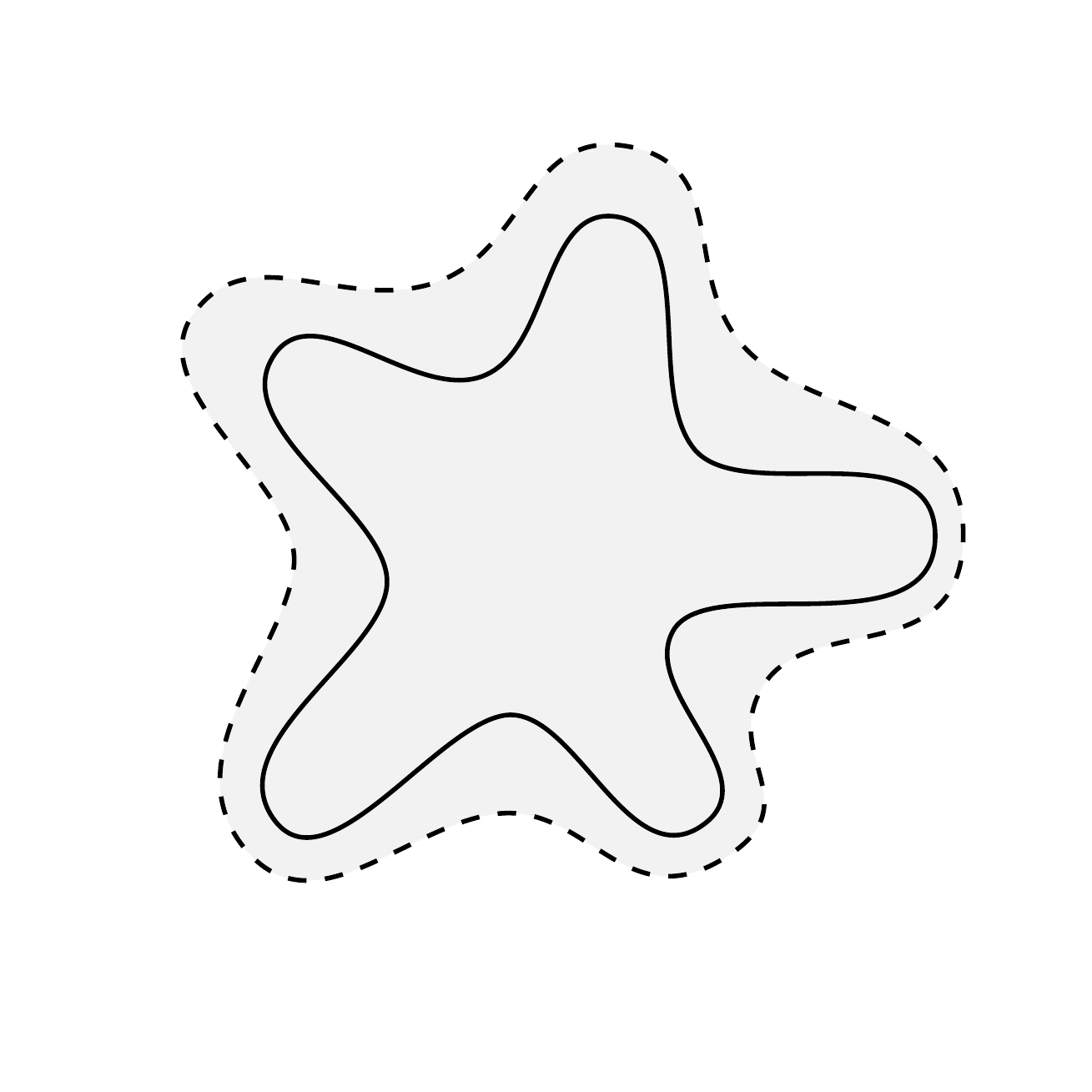}
	}
    \caption{(a) The electronic droplet in the external potential; the pointlike charges (impurities) are marked by $\otimes$. (b) The support of eigenvalues $D_e$ (solid line), and the background charge occupying the domain $D$ (dashed line).}
    \label{droplet3}
\end{figure}

A particularly important special case arises if $U(z)$ is a quasiharmonic potential $U(z)=-|z|^2+2 \Re V(z)$, where $V(z)$  is an analytic function in the domain, which includes the support of eigenvalues $D_e$, and $\Re$ denotes the real part. The following two interpretations of $U(z)$ are of equal importance: (1) $2\Re V(z)$ is the electric potential created by {\it impurities} (pointlike electric charges\footnote{Equivalently, one can consider magnetic impurities, i.e.,\ magnetic filament fluxes orthogonal to the plane.}) located in the exterior of $D_e$ [Fig.~\subref*{droplet3-a}], and (2) $U(z)$ is the potential generated by the {\it background charge}, which fills the domain $D\supset D_e$ with a unit density [Fig.~\subref*{droplet3-b}], i.e.,
\begin{equation}\label{U-def}
	U(z)=-\frac{2}{\pi}\int_D\log|z-z'|d^2z.
\end{equation}
Both in the interior and exterior of $D$, the potential $U(z)$ admits the following series expansions,
\begin{equation}\label{U}
	U(z)=\begin{cases}
	-|z|^2-V_0+2\Re\sum_{k>0}T_k z^k,& z\in D,\\
	-2T_0 \log|z|+2\Re\sum_{k>0}V_k \frac{z^{-k}}{k},& z\in \mathbb C\setminus D.
		\end{cases}
\end{equation}
The coefficients of the series expansions $T_k$ and $V_k$ are the harmonic moments of $D$, namely,
\begin{equation}\label{TV}
	\begin{gathered}
	T_k=\frac{1}{\pi k}\int_{\mathbb C\setminus D} z^{-k}d^2z,\qquad V_k=\frac{1}{\pi}\int_{D} z^kd^2z,\\
	T_0=\frac{1}{\pi}\int_{D} d^2z,\qquad V_0=\frac{2}{\pi}\int_{D} \log|z|d^2z.
	\end{gathered}
\end{equation}

The mean values of the symmetric functions of the particle coordinates are defined in the usual way,
\begin{equation}
	\langle A \rangle=\int_{-i\infty}^{i\infty}  d\mu \int_{\mathbb C^N} A(z_1,\dotsc,z_N) e^{\beta(F+\mu N-E)}\prod_{i=1}^N d^2z_i.
\end{equation}
Some important examples of the symmetric functions are the logarithmic function $v_0$ and polynomials $v_k$,
\begin{equation}\label{v-sym}
	v_0=\frac{2}{\pi}\sum_{i=1}^N\log|z_i|,\qquad v_k=\frac{1}{\pi}\sum_{i=1}^N z_i^k.
\end{equation}
In the large $N$ limit these functions become the harmonic moments of the support of eigenvalues $D_e$, defined in a similar way to~\eqref{TV}. By $t_k$ and $t_0$ we will denote the complement set of the harmonic moments of $D_e$.

\subsection{Large N limit}

At large $N$ and low enough temperature the Dyson gas behaves as an incompressible charged liquid. It is convenient to introduce the short distance cutoff $\hbar$ (which is simply the area of a cell occupied by a single electron) and define the quasiclassical limit as $N\to\infty$ and $\hbar\to 0$ while $t_0=N\hbar$ is kept fixed. Then, the eigenvalues densely fill the domain $D_e$ in a complex plane, so that the mean density $\rho(z)$  in the exterior of $D_e$ is exponentially small as $N\to\infty$. Since the density of eigenvalues is a smooth function in the large $N$ limit, the partition function~\eqref{Z} can be rewritten as
\begin{equation}\label{Z-rho}
	e^{-\beta F}=\int_{-i\infty}^{i\infty} d\mu\int e^{\frac{\beta}{\hbar^2}(\mu t_0 -  E[\rho])}[d\rho],
\end{equation}
where $t_0$ is the area of $D_e$, and we rescaled $\mu\to\mu/\hbar$ for convenience. The Jacobian of the transformation, $\prod_i d^2z_i=J[\rho][d\rho]$, is $ J[\rho]=\exp[\hbar^{-1}\int_{D_e}\rho\log\rho d^2z+cN]$, where $c$ is a constant~\cite{LargeN}. The energy of the Dyson gas~\eqref{E} can be written in terms of the density function. In the leading order it equals\footnote{The correction $\frac{\hbar}{2}(2-\beta)\int \rho\log \rho\, d^2z$, which results from the atomic structure of the Dyson gas, can be neglected at large scales.}
\begin{equation}\label{E-cont}
	- \hbar^2 E=\iint \rho(z) \log|z-z'| \rho(z') d^2zd^2z'+\int U(z)\rho(z)d^2z.
\end{equation}

Minimizing the functional $\mu t_0 -  E[\rho]$ in~\eqref{Z-rho}, we obtain the following equation for the charge distribution,
\begin{equation}\label{Extremum}
	-2\int \rho(z)\log|z-z'|d^2z'=U(z)+\mu.
\end{equation}
The classical (equilibrium) solution to this equation, which is $\rho_{cl}(z)=1/\pi$, if $z\in D_e$, and $\rho_{cl}(z)=0$ otherwise, determines the support of eigenvalues $D_e$ in what follows. All harmonic moments of the domains $D_e$ and $D$ are equal, except for $v_0\neq V_0$ (or,  equivalently, $t_0\neq T_0$) because of the term $\mu$ in~\eqref{Extremum}. If $t_0=T_0$, the domains $D_e$ and $D$ coincide [Fig.~\subref*{droplet3-b}], while in the limit $t_0\to0$ the support of the eigenvalues is the mother body of $D$~\cite{Gustafsson}.

The extremum condition~\eqref{Extremum} also fixes the Lagrange multiplier $\mu$.  Comparing the potential of the Dyson gas at the origin with $U(0)$, we obtain from~\eqref{Extremum}
\begin{equation}\label{mu-dif}
	\mu=V_0-v_0.
\end{equation}
After substituting this extremal value of $\mu$ into~\eqref{Z-rho}, the partition function of the gas takes the form
\begin{equation}\label{Frho}
	e^{-\beta F}=\int e^{\frac{\beta}{\hbar^2}(\mu t_0 - E[\rho])}[d\rho].
\end{equation}

The free energy admits the $1/N$ expansion~\cite{LargeN}, and its leading contribution to $F$ is determined by the classical configuration $\rho_{cl}(z)$. The $1/N$ corrections to the free energy (which we do not consider here) emerge as the atomic structure of the Dyson gas is taken into account~\cite{LargeCor,LargeN}. 

The language of statistical thermodynamics provides an adequate description of the Dyson gas in the large $N$ limit. One can introduce the entropy $S=\beta^2\p F/\p \beta$,\footnote{The classical configuration $\rho_{cl}(z)$ does not contribute to the entropy, as the charges are frozen at their equilibrium positions. A nontrivial contribution to the entropy results from the discrete ``atomic'' structure of the Dyson gas.} the internal energy $\langle E\rangle=F+S/\beta$, and external coordinates, which determine the configuration of the system~\cite{Gibbs}. These coordinates are the parameters of the external potential~\eqref{U}, i.e., $V_0$, $T_1$, $\overline{T}_1$, etc. By differentiating~\eqref{Frho} we obtain the {\it fundamental thermodynamic relation}, which relates changes of the entropy, the energy, and the generalized ``forces'' $\langle v_k\rangle=-\langle\p E/\p T_k\rangle $,
\begin{equation}\label{dE}
	d\langle E \rangle = \Theta dS - T_0 dV_0 + 2 \Re\sum_{k>0}V_k dT_k+\mu dt_0,
\end{equation}
where $\Theta\equiv1/\beta$ is a temperature, and we took into account the mean values of the ``forces'', $\langle v_k \rangle=V_k$ and $\langle t_0\rangle=T_0 $, in the thermodynamic equilibrium.

\subsection{The Laplacian growth equation}

The {\it adiabatic} (quasistatic) variations of the parameters of the potential $U(z)$ cause the evolution of the support of eigenvalues, that is a motion of the boundary $\p D_e$. The dynamical law which governs the evolution is incorporated in the extremum condition. By differentiating~\eqref{Extremum} with respect to $z$ and replacing $\rho_{cl}(z)$ with the characteristic function of $D_e$, we obtain the functional equation for the support of the eigenvalues $D_e$,
\begin{equation}\label{Extremum2}
	\frac{1}{\pi}\int_{D_e}\frac{d^2z'}{z-z'}=U'(z).
\end{equation}
Taking the time derivative of~\eqref{Extremum2} we arrive at the celebrated Laplacian growth equation~\cite{LGeq},
\begin{equation}\label{LG}
	\dot S(z)=2W'(z),
\end{equation}
written in terms of the Schwarz function $S(z)$ for the boundary $\p D_e$,\footnote{The Schwarz function is defined as the analytic continuation of $\bar z=S(z)$ away from the contour~\cite{Davis}.} and of the complex potential, $W(z)$, such that $\dot U(z)=W(z)$ if $z\in \mathbb C\setminus D$ and analytically continued inside $D$.

\section{Fluctuations}

\subsection{Linear thermodynamics}

Following Einstein's theory of thermodynamic fluctuations~\cite{Einstein}, the probability of the fluctuation $P\propto \exp \Delta S$ is determined by the entropy difference $\Delta S$ between two thermodynamic states. The change of the entropy of the system $\Delta S=-\beta R_{min}$ can be calculated using the minimal reversible work $R_{min}$, which has to be applied to bring the system out of equilibrium. For $R_{min}$ one can use the expression~\cite{L5}
\begin{equation}
	R_{min}=\Delta \langle E\rangle-\Theta \Delta S+T_0 \Delta V_0 - 2\Re\sum_{k>0}V_k \Delta T_k-\mu\Delta t_0 ,
\end{equation}
where $\Theta$, $T_0$, $V_k$, $t_0$ are the equilibrium values of temperature, harmonic moments, and the area of the liquid droplet. Using the fundamental thermodynamic relation~\eqref{dE} we find in the quadratic approximation the logarithm of the probability for occurring fluctuations of the thermodynamic variables,
\begin{equation}\label{logP}
	\log P\propto \frac{\Delta V_0 \Delta T_0-2\Re\sum\Delta V_k \Delta T_k -\Delta \mu \Delta t_0-\Delta S \Delta \Theta }{2\Theta}.
\end{equation}

The large $N$ limit of the Dyson gas corresponds to a very low temperature, when the charges are frozen at their equilibrium positions, and their fluctuations are negligible. Thus, we will consider the classical fluctuations at the constant temperature $\Delta \Theta=0$ only.

\subsection{Stochastic growth}

Remarkably, highly unstable and nonequilibrium Laplacian growth can be considered as small (Gaussian) fluctuations near equilibrium in the beta ensemble of complex eigenvalues. Different ways of adding new particles to the system result in different growth processes. A particularly important example is an aggregation of diffusing particles issued from the source at infinity. Since the particles can freely arrive at the system to compensate the background charge, we have $D_e=D$ and $\mu=0$ at the thermodynamics equilibrium~\eqref{Extremum}.

First, we introduce {\it the deterministic growth}, which serves as a reference point in the space of all growth processes. It is caused by a specific variation in the background potential, when all $T_k$'s (except $T_0$) are kept fixed. The new particles arrive at the system from the distant source to compensate the change of $U(z)$. Being attached to the boundary of $D_e$ with a rate proportional to the local electric field, they cause the growth of the support of eigenvalues. Since the change in $U(z)$ is a much faster process than adjusting the domain $D_e$,\footnote{The newly arrived particles undergo a diffusion from the source to the interface.} the chemical potential for newly incoming particles is nonzero and is given by~\eqref{mu-dif}. Expanding this formula in $\Delta T_0\ll T_0$, we obtain
\begin{equation}\label{mu-logR}
	\mu=2\log R\cdot \Delta T_0 + O[(\Delta T_0)^2],
\end{equation}
where $2\log R=\p V_0/\p T_0$~\cite{Mark2000}, and $R=z'(\infty)$ is the conformal radius, determined by the conformal map $z(w)$ from the exterior of the unit disk, $|w|>1$, to the exterior of $D$,\footnote{Here and below we do not indicate the time dependence explicitly and write simply $z(w)$.}
\begin{equation}\label{z-def}
	z(w)=Rw+c_0+\sum_{k>0}c_k w^{-k},\qquad w\to\infty.
\end{equation}

Representing the logarithm of the conformal radius as the solution to the Dirichlet boundary problem,\footnote{Namely, $\log R=-(1/2\pi)\oint_{\p D}\log|w'(\zeta)|\p_n G(\zeta,\infty)|d\zeta|$, where $G(z,z')$ is the Green's function of the Dirichlet boundary problem in the exterior of $D$~\cite{Schiffer}, and $\p_nG(\zeta,\infty)=-|w'(\zeta)|$ at the boundary.} we recast the chemical potential~\eqref{mu-logR} in the form
\begin{equation}\label{mu-l}
	\mu=-2\Delta t\oint_{\p D}u_n(\zeta)\log|w'(\zeta)| |d\zeta|,
\end{equation}
where $w(z)$ is the inverse to the conformal map~\eqref{z-def}, and $u_n(\zeta)\Delta t=|w'(\zeta)|(\Delta T_0/2\pi)$ is the normal displacement of the boundary $\p D$ due to the change $\Delta T_0$. The newly attached particles cause the growth of $D_e$ with the normal velocity $u_n(\zeta) = (q/2\pi)|w'(\zeta)|$ which prompts Darcy's law, where $q=\Delta T_0/\Delta t$ is the rate of the growth. Thus, the evolution of the interface follows the {\it deterministic Laplacian growth dynamics},
\begin{equation}\label{LG-inf}
	\Im(\p_t\bar z \p_\phi z)=\frac{q}{2\pi},
\end{equation}
where $\phi=-i\log w(\zeta)$ parametrizes the boundary, $\Im$ denotes the imaginary part, and the subscripts stand for partial derivatives. By plugging $\Delta T_k=0$, $\Delta t_0=\Delta T_0$, and $\Delta V_0=\Delta \mu$ into~\eqref{logP},  we conclude that deterministic Laplacian growth~\eqref{LG-inf} maximizes the probability of fluctuations.

{\it The stochastic growth processes} deviate from the deterministic Laplacian growth dynamics~\eqref{LG-inf} by random fluctuations of the interface during the evolution. These fluctuations are generated by newly arrived particles, which tend to compensate the change of the parameters of the external potential $T_k\ (k\geq0)$ caused by variations of the impurities [Fig.~\subref*{droplet3-a}]. The latter are related by the Richardson theorem~\cite{Richardson},
\begin{equation}\label{DT}
	\Delta T_0=\frac{1}{\pi}\sum_{m=1}^M q_m,\quad\Delta T_k=\frac{\Delta t}{\pi}\sum_{m=1}^M q_m \frac{A_m^{-k}}{k},
\end{equation}
where $A_m$ is the position of $m$th impurity, and $q_m$ is its rate of change of the charge. The corresponding interface dynamics is then described by the {\it stochastic Laplacian growth equation},
\begin{equation}\label{LG-stochastic}
	\Im(\p_t\bar z \p_\phi z)=\frac{q}{2\pi}+\sum_{m=1}^M\frac{q_m}{2\pi}\Re\frac{e^{i\phi}+a_m}{e^{i\phi}-a_m},
\end{equation}
where $a_m$ are the inverse preimages of $A_m=f(1/\bar a_m)$ inside the unit disk. This equation is well known in the classical Laplacian growth problem in the presence of pointlike hydrodynamical sources $q_m$ at the points $A_m$~\cite{Kufarev}.

The general formula~\eqref{logP} determines small fluctuations of all basic thermodynamic quantities. The harmonic moments $T_k$ and $V_k$, as the coefficients of the multipole expansion of the potential~\eqref{U-def}, are interrelated via a certain potential function~\cite{Mark2000},
\begin{equation}\label{TV-tau}
	V_k=\frac{\p \log \tau}{\p T_k},\quad \overline V_k=\frac{\p \log \tau}{\p \overline T_k},\quad V_0=\frac{\p \log \tau}{\p T_0},
\end{equation}
where $\log \tau$ is the logarithm of the tau function of the boundary curve~\cite{tau2000},
\begin{equation}\label{tau-def}
	\log \tau=-\frac{1}{\pi^2}\int_D\int_D \log\left|\frac{1}{z}-\frac{1}{z'}\right|d^2zd^2z'.
\end{equation}
The symmetry relations between the harmonic moments suggest to chose $T_k$'s as independent variables.

Variations of $T_k$'s also cause the deviation of the chemical potential for newly incoming particles from its ``deterministic'' value~\eqref{mu-l}. Since $u_n(\zeta)|d\zeta|=\dot S(\zeta)d\zeta/2i$, we transform the line integral in~\eqref{mu-l} in the contour one. The Laplacian growth equation~\eqref{LG} relates the change of the Schwarz function with the variation of the electric potential, $\pi\dot S^+(z)=-\sum q_m/(z-A_m)$.\footnote{Here $S^+(z)$ is a regular in $D$ part of the Schwarz function $S(z)=S^+(z)+S^-(z)$.} Thus, by evaluating the contour integrals, we determine the variation of the chemical potential,
\begin{equation}\label{D-mu}
	\Delta \mu = - 2\Delta t\sum_{m=1}^M q_m \log|w'(A_m)|.
\end{equation}

Now, using the symmetry relation~\eqref{TV-tau} one can express $\Delta V_k$'s in terms of $\Delta T_k$'s to transform the probability for occurring fluctuation~\eqref{logP} in the quadratic form in $\Delta T_k$ and $\Delta \overline T_k$ only. As for the term $\Delta t_0\Delta \mu$,\footnote{Recall that $\Delta t_0=\Delta T_0$.} we use~\eqref{D-mu} and take into account the description of the conformal map, $w(z)$, in terms of the tau-function, which leads to the identity~\cite{tau2000}
\begin{equation}\label{w'}
	\log|w'(A)|=-\frac{\p^2\log \tau}{\p T_0^2}+2\Re\sum_{k,l>0}\frac{A^{-k-l}}{kl}\frac{\p^2\log \tau}{\p T_k \p T_l},
\end{equation}
thus allowing one to express $\Delta \mu$ through $\Delta T_k$'s~\eqref{DT}. As discussed earlier, the variations in $T_k$'s are caused by fluctuations of the pointlike impurities in the exterior of $D_e$~\eqref{DT} [Fig.~\subref*{droplet3-a}]. Thus, the quadratic form in the exponent of~\eqref{logP} (upon rewriting in terms of $T_k$'s) includes the cross terms of the form $c_{mn}q_mq_n$. These terms, however, do not contribute to the mean fluctuation values of any quantities, if the statistical independence of impurities, $\langle q_m q_n\rangle=0$ for $m\neq n$, is assumed. Under these circumstances the probability for occurring fluctuations can be written in the form\footnote{Equivalently, one may think that only a single impurity fluctuates per time unit.}
\begin{equation}\label{P-l-tt}
	P\propto \exp\left\{ -\frac{\beta}{\hbar^2} \sum_{k,l>0} \frac{\p^2 \log \tau}{\p T_k\p \overline T_l}\Delta T_k \Delta \overline T_l \right\}.
\end{equation}
The mean square fluctuations of the harmonic moments,
\begin{equation}\label{MTT}
	\langle \Delta T_k \Delta \overline T_l\rangle=\beta^{-1}\hbar^2g_{k\bar l},
\end{equation}
then are determined by the coefficients of the quadratic form in the exponent of~\eqref{P-l-tt},
\begin{equation}\label{gkl}
	g^{k\bar l}(T_0,T_1,\overline T_1,\dotsc)=\frac{\p^2 \log \tau}{\p T_k\p \overline T_l},
\end{equation}
and $g_{k\bar l}$ are the elements of the inverse matrix. The mean square fluctuations of $\Delta V_k$'s are obtained from~\eqref{MTT} by virtue of the symmetry relation~\eqref{TV-tau}. It also follows from~\eqref{P-l-tt} that the area of the droplet does not fluctuate with time, as we consider fluctuations of the shape of the interface against the background of the growth process with a constant rate.

\subsection{Path-integral formulation of the Laplacian growth}

If growth continues until time $T\gg \Delta t$ ($T/\Delta t$ is an integer) the associated probability of the process is given by the product of the {\it conditional probabilities}~\eqref{P-l-tt},
\begin{equation}\label{Prod}
	\mathcal P=P(t_1)P(t_2)\cdots P(t_{T/\Delta t}).
\end{equation}
Since $P(t_{i+1})$ depends solely (through the tau function) on the present domain at time~$t_i$, the product in~\eqref{Prod} is a Markovian chain. In the limit $\Delta t\to 0$ by summing the exponents of the probabilities, we obtain
\begin{equation}\label{Paction}
	\mathcal P \propto\exp\left\{ -\frac{\beta \Delta t	}{\hbar^2}\int_0^T dt \sum_{k,l>0}g^{k\bar l}\frac{dT_k}{dt}\frac{d\overline{T_l}}{dt}  \right\}.
\end{equation}

Remarkably, the exponent in~\eqref{Paction} is the action for a nonrelativistic massless particle on the infinite-dimensional complex manifold with the K\"ahler metric~\eqref{gkl}, so $\log \tau$ is the K\"ahler potential~\cite{Takhtajan}.

Although the growth of the support of eigenvalues was treated as response to the variation of the external potential, one can forget the details of this derivation and consider the growth probability on its own. Then, assuming that fluctuations of the interface during evolution are inevitable features of unstable Laplacian growth, formula~\eqref{Paction} allows to compare probabilities of different growth scenarios, which lead to the same final domain starting, e.g., from the ideal circle. Under these circumstances, the path-integral formulation does what it is supposed to do, namely, indicates the most probable growth scenario.

\section{Conclusion and discussion}

The application of statistical {\it equilibrium} mechanics to stochastic growth phenomena sheds light on the long-standing problems of {\it nonequilibrium} universal growth. In this paper we have touched only on the Laplacian growth, because of its impressively wide applicability ranging from solidification and oil recovery to biological growth.

A particularly important result is an explicit calculation of the growth probability~\eqref{Paction}, which is simply the propagator of a free particle on the infinite-dimensional complex manifold with the K\"ahler metric. From the Lagrangian mechanics point of view, geodesics in this space determine the most probable growth scenarios of 2D planar domains. Therefore, the pattern selection problem in Laplacian growth might be considered in the purely classical mechanics framework.

The fluctuation theory of the Laplacian growth also has a rich mathematical structure if one uses its dual description (known as the inverse potential problem) in terms of the conformal maps. The growth probability~\eqref{P-l-tt} can be obtained also by considering the ``entropy of the layer'' that is composed of particles which attach to the boundary of the domain per time unit~\cite{AM1,AM2}. This observation allows to relate the Laplacian growth problem to the theory of random partitions~\cite{Okounkov}, and might have far reaching consequences.

For direct applications in physics, our results assume the following impact. The probability~\eqref{P-l-tt} generalizes the growth process of the electronic droplet in the quantum Hall regime~\cite{VF} to the case when a large number of electrons attach to the boundary of the droplet during a single time unit. It allows one to determine the correlation functions between the electrons at the boundary of the droplet, and consider the geometrical properties of the boundary, e.g., its multi-fractal spectrum.

The Gibbs-Boltzmann statistics of fluctuations~\eqref{P-l-tt} provides a framework for systematic {\it analytic} treatment of static (time-independent) perturbations of highly nonequilibrium growth processes, which has been lacking so far. It also furnishes a clue to the {\it dynamical} (time-dependent) fluctuations, which reveal the role of the fluctuation-dissipation theorem in the Laplacian growth problem. The latter problem is closely related to the probability distribution function for fluctuating impurities, which appear in the right-hand side of the stochastic Laplacian growth equation~\eqref{LG-stochastic}. The time dependent fluctuations promises to elucidate derivation of the Laplacian growth fractal spectrum, and unexplained selections of patterns. These problems will be addressed in future publications.

\bibliography{biblio}{}

\end{document}